\documentclass[12pt,a4paper]{article}
\usepackage[text={16.5cm,23cm},centering]{geometry}
\usepackage{amsmath}
\usepackage{amsthm}
\usepackage[utf8]{inputenc}
\usepackage{tgtermes,tgheros,tgcursor}
\usepackage[varg]{newtxmath}
\usepackage{hyperref}
\hypersetup{%
 colorlinks=true,%
 linkcolor=blue,
 citecolor=blue,
}
\usepackage{xspace}

\numberwithin{equation}{section}

\newcommand{\del}{\partial}
\newcommand{\Ncal}{\mathcal{N}}
\newcommand{\Zb}{\mathbb{Z}}
\DeclareMathOperator*{\Tr}{\mathrm{Tr}}

\newcommand{\slz}{SL(2, $\Zb$)\xspace}
\newcommand{\xyz}{\#}
\newcommand{\deltaB}{\boldsymbol{\delta}_{\mathrm{B}}}

\newcommand{\Cs}{C^{(4,y_{\xyz})}}
\newcommand{\Co}{C^{(1,y_2,y_3)}}

\newcommand{\Lcal}{\mathcal{L}}
\newcommand{\phit}{\tilde{\phi}}
\newcommand{\At}{\widetilde{A}}
\newcommand{\Bt}{\widetilde{B}}

\newcommand{\Iphi}{I_{\phi}}

\begin{document}
\begin{center}
  \begin{flushright}
    OU-HET 1153\\
    YITP-22-86
  \end{flushright}
  \vspace{8ex}
  {\Large \bfseries \boldmath SL$(2,\mathbb{Z})$ action on quantum field theories with U(1) subsystem symmetry}\\
  \vspace{4ex}
  {\Large Satoshi Yamaguchi}\\
  \vspace{2ex}
  {\itshape Department of Physics, Graduate School of Science, 
  \\
  Osaka University, Toyonaka, Osaka 560-0043, Japan}\\
  {\centering and}
  \\  
  {\itshape  Yukawa Institute for Theoretical Physics, Kyoto University  
  \\
  Kitashirakawa-Oiwakecho, Sakyo-Ku, Kyoto 606-8502, Japan}\\
  \vspace{1ex}
  \texttt{yamaguch@het.phys.sci.osaka-u.ac.jp}\\
  \begin{abstract}
    We consider SL$(2,\mathbb{Z})$ action on quantum field theories with U(1) subsystem symmetry in five dimensions.  This is an analog of the SL$(2,\mathbb{Z})$ action considered in \cite{Witten:2003ya}.  We show that the exotic level 1 BF theory and the exotic level 1 Chern-Simons theories are trivial and almost trivial, respectively.  By using this fact, we define S operation and T operation.  These operations make SL$(2,\mathbb{Z})$ group up to a possible invertible phase that is unity within the space-times treated in this paper.  We also demonstrate SL$(2,\mathbb{Z})$ action on the $\varphi$ theory as an example.
  \end{abstract}
\end{center}

\vspace{4ex}
\section{Introduction and summary}

Subsystem symmetries are a class of exotic symmetries in non-relativistic quantum field theories \cite{Batista:2004sc,Nussinov:2006iva,Nussinov:2009zz} (See \cite{Cordova:2022ruw} and references therein for recent developments).
In particular, they play an important role in fracton phases \cite{Chamon:2004lew,Haah:2011drr,Nandkishore:2018sel,Pretko:2020cko}.
Such exotic systems have been mainly studied in soluble lattice quantum mechanics \cite{Vijay:2015mka,Vijay:2016phm}. They have also been studied by other approaches such as foliated quantum field theory \cite{Slagle:2020ugk,Hsin:2021mjn},  infinite component Chern-Simons theory \cite{ma2020fractonic}, and  continuous quantum field theory \cite{Pretko:2016kxt,Pretko:2016lgv,2018PhRvB..98c5111M,Bulmash:2018lid,Seiberg:2019vrp,Seiberg:2020bhn,Seiberg:2020wsg,Seiberg:2020cxy,Gorantla:2020xap}.
In this paper, we employ continuous quantum field theory approach.
Subsystem symmetries and fracton phases have also been investigated in the interests of high-energy physics, such as supersymmetry \cite{Yamaguchi:2021qrx,Katsura:2022xkg}, quiver gauge theory \cite{Razamat:2021jkx,Franco:2022ziy}, and string theory \cite{Geng:2021cmq}.

Another important ingredient of this paper is \slz action on the 3D quantum field theories (QFTs) with U(1) global symmetry\cite{Burgess:2000kj,Witten:2003ya}.  
This is not a symmetry nor a duality in general.  This produces an \slz orbit of QFTs that are, in general, inequivalent theories.
From the holographic point of view, all those theories related by \slz transformation correspond to the same bulk theory with different boundary conditions \cite{Witten:2003ya}.  
Even if non-holographic case, if the partition function of the 3D theory can be interpreted as the wave function of a 4D theory, the partition functions related by \slz correspond to the same state, but the basis are different.
\slz action on QFTs with discrete higher-form symmetry is also proposed by \cite{Gaiotto:2014kfa,Bhardwaj:2020ymp}. 
Recently, such \slz action provide a nice way to find non-invertible symmetries \cite{Choi:2021kmx,Kaidi:2021xfk} (see also \cite{Koide:2021zxj}).  Another interesting direction is the generalization to the higher spin symmetry studied in \cite{Leigh:2003ez}.

In this paper, we propose an \slz action on 5D QFTs with U(1) subsystem symmetry.  
We first show that the level 1 5D exotic BF theory is a trivial theory.  Then we define S operation and T operation.  
We show that they form the \slz group.  
We also consider the $\phi$ theory \cite{Seiberg:2020bhn,Gorantla:2020xap} as an example and \slz action on it.

It will be an interesting future problem to investigate \slz action on QFTs with discrete (higher-form) subsystem symmetry.  In some case, we may find non-invertible subsystem symmetry by using this \slz action.

The construction of this paper is as follows.  In Sec.~\ref{sec:BF}, we investigate 5-dimensional BF theory at level 1.  We will show that it is a trivial theory. In Sec.~\ref{sec:ST}, we define S and T operation, and show that they generate \slz .  In Sec.~\ref{sec:phi}, we choose the $\phi$ theory as an example, and calculate \slz transformation of the $\phi$ theory.

\section{Exotic BF theory and exotic Chern-Simons theory}
\label{sec:BF}
In this section, we study the exotic BF theory and the exotic Chern-Simons theory at level 1 as preliminary.  Such exotic BF theories and Chern-Simons theories have been investigated in \cite{you2020fractonic,Burnell:2021reh,Yamaguchi:2021xeq}.

\subsection{Subsystem symmetry}
\label{sec:subsystemsymmetry}
We mostly consider Euclidean spacetime formalism in this paper.
Let the Euclidean space-time coordinates $x_{\mu},\ \mu=1,2,3,4,5$. For simplicity, we focus on $T^5$ obtained by identifying $x_{\mu}\sim x_{\mu}+\ell_{\mu}$.

We introduce a symbol $\xyz$, and define the derivative $\del_{\xyz}$ by
\begin{align}
  \del_{\xyz}:=\del_{1}\del_{2}\del_{3}.
\end{align}
We use labels $a,b,\dots=\xyz,4,5$, and they are raised and lowered by $\delta^{ab}$ and $\delta_{ab}$, respectively.  Then, for example, a differential operator analogous to the Laplacian can be written as
\begin{align}
  \del_{a}\del^{a}=\del_{\xyz}^{2}+\del_{4}^2+\del_{5}^2=(\del_{1}\del_{2}\del_{3})^2+\del_{4}^2+\del_{5}^2.
\end{align}
The following Leibniz rule is useful:
\begin{align}
  (\del_{a}A)B+A(\del_{a}B)=(\text{total derivative}),
\end{align}
where $A,B$ are functions on this spacetime.  This equation allows us to perform the integration by parts.

The three-torus coordinated by $x_1,x_2,x_3$ is denoted by $T^{3}_{\xyz}$, while two-torus coordinated by $x_4,x_5$ is denoted by $T^2_{45}$.  We use the notation $x_{\xyz}=(x_1,x_2,x_3)$ for shorthand.

In the discussion in this paper, we often need a cut-off in $T^3_{\xyz}$.  In such case, we employ the following lattice cutoff.  Let $a$ be the lattice spacing and $L_i=\ell_i/a \in \Zb,\ (i=1,2,3)$ be the number of lattice sites on the $i$-axis. We define the set of lattice sites as
\begin{align}
  \Omega=\{
    y_{\xyz}\in T^3_{\xyz}| y_i \in a \Zb,\ i=1,2,3\}.
\end{align}
We also define the set of lattice sites on the $i$-axis
\begin{align}
  \Omega_{i}=a\Zb/\ell_i\Zb,
\end{align}
and the $ij$-plane
\begin{align}
  \Omega_{ij}=\Omega_{i}\times \Omega_{j}.\label{ijlatticesites}
\end{align}

The current $J^a,\ a=\xyz,4,5$ for the subsystem symmetry that we consider in this paper satisfies
\begin{align}
  \del_{a}J^{a}=0.
  \label{conservationlaw}
\end{align}
We can construct huge number of conserved charges from this current as follows.  Let us choose a three-torus $T^3_{145}(x_2,x_3)\subset T^3_{\xyz}\times T^{2}_{45}$ obtained by fixing $x_2,x_3$ to be constant values.  We consider two-form $K^{1}$ defined as
\begin{align}
  K^{1}&=
  \del_2\del_3 J^{\xyz}dx^4\wedge dx^5+  J^{4}dx^5\wedge dx^1+J^5 dx^1\wedge dx^4.
\end{align}
This $K^{1}$ satisfies
\begin{align}
  dK^{1}|_{T^3_{145}(x_2,x_3)}=0,
\end{align}
due to \eqref{conservationlaw}. 
Thus, for an arbitrary closed surface $\Sigma \subset T^3_{145}(x_2,x_3)$, we obtain a ``conserved charge''
\begin{align}
  Q^{1,(x_2,x_3)}(\Sigma):=-i\int_{\Sigma} K^{1}.
\end{align}
This $Q^{1,(x_2,x_3)}(\Sigma)$ is invariant under the continuous deformation of $\Sigma$ within $T^3_{145}(x_2,x_3)$ unless it crosses another operator.
There are other conserved charges $Q^{2,(x_3,x_1)}(\Sigma),\ Q^{3,(x_1,x_2)}(\Sigma)$ defined as the same way.  They are not completely independent, but satisfy certain constraints \cite{Seiberg:2020bhn,Gorantla:2020xap}.

Let us introduce a gauge field  $A_{a}(x),\ a=\xyz,4,5$ for this U(1) subsystem symmetry.  It couples to $J^{a}(x)$ as
\begin{align}
  \int d^5 x A_{a}(x)J^{a}(x).\label{coupling}
\end{align}
The gauge transformation is given by
\begin{align}
  A_{a}\to A'_{a}=A_{a}+\del_{a}\lambda.
\end{align}
$\lambda(x)$ is a bosonic field which has the periodicity $\lambda\sim \lambda+2\pi$ and allows the same discontinuity as the $\phi$ field of the $\phi$ theory \cite{Seiberg:2020bhn,Gorantla:2020xap}.  Then, the coupling to the current is gauge invariant for an infinitesimal gauge transformation due to the conservation law \eqref{conservationlaw}.

The field strength is defined as
\begin{align}
  F_{ab}=\del_{a} A_{b}-\del_{b} A_{a}.
\end{align}
This field strength is gauge invariant.

It is important to notice that if a theory contains this gauge field, the theory always has the U(1) global subsystem symmetry.\footnote{Precisely speaking, if the theory contains dynamical monopoles, it does not have this U(1) global subsystem symmetry.}  The current is expressed as
\begin{align}
  M^a=\frac{i}{2\pi}\epsilon^{abc}\del_{b}A_{c},
  \label{magneticcentersymmetry}
\end{align}
where $\epsilon^{abc}$ is a completely antisymmetric symbol with $\epsilon^{\xyz 4 5}=+1$.  This current satisfies the conservation law $\del_{a}M^{a}=0$, due to the Bianchi identity.

\subsection{Delta functional for the gauge field}

In order to introduce \slz action in this paper, an analog of the ``trivial theory'' in \cite{Witten:2003ya} is quite important.
We study this theory here.  Unfortunately, we do not have sophisticated tools such as homology, cohomology, differential forms and so on for subsystem symmetries, so far.
Thus, we employ the naive component-wise approach here.

Let $A_a, B_a$ gauge fields for U(1) subsystem symmetry.  We claim that the following equation holds.
\begin{align}
  \int DB \exp\left(
    \frac{i}{2\pi}\int d^5 x 
    \epsilon^{abc}B_{a}\del_{b}A_{c}
  \right)
  =\delta[A]. \label{keyidentity}
\end{align}
Here, the path integral $\int DB$ is appropriately divided by the gauge volume, so that the partition function agrees with that of the canonical formalism.
The right-hand side is the delta functional for $A_{a}$.  $\delta[A]=0$ unless $A_{a}$ is a pure gauge.  It is normalized as
\begin{align}
  \int DA \delta[A]=1.
\end{align}

We give some arguments that support the identity \eqref{keyidentity}.

\subsubsection{Vanishing of all the Wilson loops}
Here we will show that the left-hand side of \eqref{keyidentity} vanishes unless $A_a$ is a pure gauge.

It may be easy to find that the left-hand side vanishes unless $F_{ab}:=\del_{a}A_{b}-\del_{b}A_{a}=0$ by naively integrating $B_{a}(x)$ out.
In other words $A_{a}$ must be a ``flat'' gauge field.  However, ``flatness'' does not imply pure gauge, since there may be non-trivial values of the Wilson loops. 

In order to argue that such Wilson loops must also vanish, we have to take topologically non-trivial configurations of $B_{a}$ into account.  Actually, we have a huge number of different topological charges due to the exotic nature of the subsystem symmetry as we will see.  For example, a topologically non-trivial configuration of $B_{a}$ satisfies the relation
\begin{align}
  \frac{1}{2\pi}\int dx_1dx_4(\del_{\xyz}B_{4}-\del_{4} B_{\xyz})=\sum_{(y_2,y_3)\in \Omega_{23}}
  n_{1,(y_2,y_3)}\delta(x_2-y_2)\delta(x_3-y_3),
\end{align}
where $\Omega_{23}$ is a set of lattice sites \eqref{ijlatticesites} in $23$-plane and $n_{1,(y_2,y_3)}$ are integers.  In the strict continuum limit, the species of such topological charges are infinite.  Here we assume that our theory has a nice cut-off and the number of such charges are finite.

Let $I$ be the label of the topological charge, $n_I$ be the associated topological charge, and $C^{(I)}_{a}(x)$ be the representative of the configurations with topological charge $n_I=1$.  Then, the most general form of $B_{a}$ can be written as
\begin{align}
  B_a(x)=\sum_{i}n_{I}C^{(I)}_{a}(x)+b_{a}(x),
\end{align}
where $b_{a}(x)$ is a topologically trivial configuration.  By this expression, the integration for $B_a$ fields can be written as
\begin{align}
  \int DB=\int Db \prod_{I}\sum_{n_I\in \Zb}.
\end{align}
Moreover, since the action is linear in $B_a$ we can discuss each integral and summation independently.

First, the integration over $b_a$ vanishes unless $A_{a}$ is flat.   Thus, we have only to treat a flat $A_{a}$ field configuration in the following arguments.

Second, we argue the ``magnetic flux'' in $1234$-direction.  Let us consider the basis $\Cs_{a}$ parametrized by $y_{\xyz}=(y_1,y_2,y_3)$:
\begin{align}
  \Cs_{5}&=0,\quad \Cs_{4}=0,\quad 
  \Cs_{\xyz}=-\del_{\xyz}f(x_{\xyz};y_{\xyz})\frac{x_4}{\ell_4},\\
  f(x_{\xyz};y_{\xyz})
  =&2\pi \Biggl[
    \frac{x_1x_2x_3}{\ell_1\ell_2\ell_3}
    -\frac{x_1x_2}{\ell_1\ell_2}\theta(x_3-y_3)
    -\frac{x_2x_3}{\ell_2\ell_3}\theta(x_1-y_1)
    -\frac{x_3x_1}{\ell_3\ell_1}\theta(x_2-y_2)\nonumber\\
    &+\frac{x_1}{\ell_1}\theta(x_2-y_2)\theta(x_3-y_3)
    +\frac{x_2}{\ell_2}\theta(x_3-y_3)\theta(x_1-y_1)
    +\frac{x_3}{\ell_3}\theta(x_1-y_1)\theta(x_2-y_2)
    \Biggr].\label{unitwindingphi}
\end{align}
This $f$ is a winding $\phi$ field configuration \cite{Gorantla:2020xap}.  Here $\theta(x)$ is the Heaviside step function. The derivatives $\del_{a}$ acting on $f$ here and below are always $x$ derivatives and not $y$ derivatives unless otherwise noted.
$\del_{\xyz}f$ is expressed as
\begin{align}
  \del_{\xyz}f(x_{\xyz};y_{\xyz})
  =&2\pi \Biggl[
    \frac{1}{\ell_1\ell_2\ell_3}
    -\frac{1}{\ell_1\ell_2}\delta(x_3-y_3)
    -\frac{1}{\ell_2\ell_3}\delta(x_1-y_1)
    -\frac{1}{\ell_3\ell_1}\delta(x_2-y_2)\nonumber\\
    &+\frac{1}{\ell_1}\delta(x_2-y_2)\delta(x_3-y_3)
    +\frac{1}{\ell_2}\delta(x_3-y_3)\delta(x_1-y_1)
    +\frac{1}{\ell_3}\delta(x_1-y_1)\delta(x_2-y_2)
    \Biggr].
\end{align}

The path-integral in the left-hand side of \eqref{keyidentity} includes the factor
\begin{align}
  \sum_{n_{(4,y_{\xyz})} \in \Zb} \exp(n_{(4,y_{\xyz})}I_{(4,y_{\xyz})}), \label{xyz4factor}
\end{align}
where $I_{(4,y_{\xyz})}$ is calculated as
\begin{align}
  I_{(4,y_{\xyz})}&=\frac{i}{2\pi}\int d^5 x \epsilon^{abc}\Cs_{a}\del_{b}A_{c}=
  \frac{i}{2\pi}\int d^5 x \epsilon^{abc}\del_{a} \Cs_{b}A_{c}\nonumber\\
  &=\frac{i}{2\pi}\int d^5 x (\del_{\xyz}\Cs_{4}-\del_{4}\Cs_{\xyz})A_{5}
  =\frac{i}{2\pi}\int d^3 x_{\xyz}dx_{4} (\del_{\xyz}\Cs_{4}-\del_{4}\Cs_{\xyz})w_{5}\nonumber\\
  &=\frac{i}{2\pi}\int d^3 x_{\xyz}dx_{4} \frac{1}{\ell_4} \del_{\xyz} f(x_{\xyz};y_{\xyz})w_{5},\label{tempI4y}\\
  w_5&:=\int dx_5 A_5.
\end{align}
This $w_5$ has the periodicity $w_5\sim w_5+2\pi$ by a large gauge transformation.

Since the gauge field $A_a$ is flat, $w_5$ turns out to be independent of $x_4$ by the usual argument.    However, $w_5$ may depend on $x_{\xyz}$.  Let us look closely at the $x_{\xyz}$ dependence of $w_5$.  The flatness implies
\begin{align}
  \del_{5}A_{\xyz}=\del_{\xyz}A_{5}.
\end{align}
Integrating both sides in $x_5$, we find
\begin{align}
  \del_{\xyz}w_5:=\del_{1}\del_{2}\del_{3}w_5=0.
\end{align}
General solution for this equation is 
\begin{align}
  w_{5}(x_{\xyz})
  =w_{5,1}(x_2,x_3)+w_{5,2}(x_3,x_1)+w_{5,3}(x_1,x_2).
\end{align}
The contribution from $w_{5,1}$ to $I_{(4,y_{\xyz})}$ in \eqref{tempI4y} is calculated as
\begin{align}
  \frac{i}{2\pi}\int d^3 x_{\xyz}dx_{4} \frac{1}{\ell_4}\del_{\xyz}f \,w_{5,1}(x_2,x_3)
  =iw_{5,1}(y_2,y_3).
\end{align}
Therefore, combined with the contributions from $w_{5,2}$ and $w_{5,3}$, $I_{(4,y_{\xyz})}$ is given by
\begin{align}
  I_{(4,y_{\xyz})}=iw_{5}(y_{\xyz}).
\end{align}
The summation of $n_{(4,y_{\xyz})}$ is written as
\begin{align}
  \sum_{n_{(4,y_{\xyz})}\in \Zb}\exp(in_{(4,y_{\xyz})} w_5(y_{\xyz}))=2\pi \delta(w_5(y_{\xyz})).
\end{align}
Thus, the partition function vanishes unless $w_5(y_{\xyz})=0$.  

Similarly, we can also argue that the partition function vanishes unless $\int dx_4 A_4=0\ \mod 2\pi$ in the same way.

Third, we argue the magnetic flux in $45$-direction.  This magnetic flux is
\begin{align}
  m(x_{\xyz})=\frac{1}{2\pi}\int dx_4 dx_5 G_{45},\\
  G_{ab}:=\del_{a}B_{b}-\del_{b}B_{a}.
\end{align}
This magnetic flux $m(x_{\xyz})$ is integer valued.  Since our gauge field is not a usual one, the magnetic flux may depend on $x_{\xyz}$.  Let us see how it depends.  From the Bianchi identity, we find that the magnetic flux satisfy the differential equation:
\begin{align}
  \del_{\xyz}m(x_{\xyz})
  =\frac{1}{2\pi}\int dx_4 dx_5 \del_{\xyz}G_{45}
  =\frac{1}{2\pi}\int dx_4 dx_5 (\del_{4}G_{\xyz 5}-\del_{5}G_{\xyz 4})
  =0.
\end{align}
The general solution for this equation is
\begin{align}
  m(x_{\xyz})=m_{1}(x_2,x_3)+m_{2}(x_3,x_1)+m_{3}(x_1,x_2),
\end{align}
where $m_{i}(x_j,x_k),\ i,j,k=1,2,3$ are arbitrary integer valued functions of $x_j,x_k$.

From the above observation, the basis of such gauge field can be chosen as follows.  Let us pick up $(y_2,y_3)\in \Omega_{23}$ and define the very thin tubular region as
\begin{align}
  R_{1}(y_2,y_3)=\{
  x_{\xyz}\in T^{3}_{\xyz}|
  y_2-a/2 < x_2 < y_2+a/2,\   
  y_3-a/2 < x_3 < y_3+a/2
  \},
\end{align}
where $a$ is the cutoff.
We define the characteristic function $g_{1}(x_{\xyz}, y_{\xyz})$ for $R_{1}(y_2,y_3)$:
\begin{align}
  g_{1}(x_{\xyz}; y_{\xyz})=
  \begin{cases}
    1& (x_{\xyz}\in R_{1}(y_2,y_3))\\
    0& (\text{otherwise})
  \end{cases}.
\end{align} 
Notice that this $g_{1}$ satisfies $\del_{1}g_{1}=0$ and therefore $\del_{\xyz}g_{1}=0$.  We choose basis $\Co_{a}$ as
\begin{align}
  \Co_{\xyz}=\Co_{4}=0,\qquad
  \Co_{5}=2\pi g_{1}(x_{\xyz}; y_{\xyz})\frac{x_4}{\ell_4 \ell_5}.
\end{align}
The magnetic flux for this basis is $m_{1}(x_2,x_3)=g_{1}(x_{\xyz};y_{\xyz}),\ m_2=m_3=0$.

The path-integral in the left-hand side of \eqref{keyidentity} includes the factor
\begin{align}
  &\sum_{n_{(1,y_2,y_3)}\in \Zb}
  \exp(n_{(1,y_2,y_3)}I_{(1,y_2,y_3)}),\label{factor1y2y30}\\
  &I_{(1,y_2,y_3)}:=\frac{i}{2\pi} \int d^5x \epsilon^{abc}\Co_{a}\del_{b}A_{c}.\label{factor1y2y3}
\end{align}
Let us calculate the coefficient $I_{(1,y_2,y_3)}$:
\begin{align}
  I_{(1,y_2,y_3)}=&i\int d^5 x \frac{1}{\ell_4 \ell_5}g_{1}(x_{\xyz};y_{\xyz})A_{\xyz}
  =i\int dx_2 dx_3 g_{1}(x_{\xyz};y_{\xyz})w_{1}(x_2,x_3)\nonumber\\
  =:&i W_1(y_2,y_3),\\
  w_{1}(x_2,x_3):=&\int dx_{1} A_{\xyz}(x).
\end{align}
Here we use the fact that the ``Wilson loop'' $w_1$ is independent of $x_4,x_5$ since $A_{a}$ is flat. This $W_1(y_2,y_3)$ is smeared Wilson loop $w_1$ in the tubular region $R_1(y_2,y_3)$.  In this smearing process, we lose finer information than the cut-off $a$.

Let us see the periodicity of these Wilson loops.
Since $f$ in Eq.~\eqref{unitwindingphi} is a nice $\phi$ field configuration and therefore a nice gauge transformation parameter, 
we obtain the gauge equivalence of the gauge fields:
\begin{align}
  A_{\xyz}\sim A_{\xyz}+\del_{\xyz}f \ \Rightarrow \ 
  w_1(x_2,x_3)\sim w_1(x_2,x_3) + 2\pi\delta(x_2-y_2)\delta(x_3-y_3).
\end{align}
As a result, we obtain the periodicity
\begin{align}  
  W_1(y_2,y_3)\sim W_1(y_2,y_3)+2\pi.
\end{align}
The factor \eqref{factor1y2y30} becomes
\begin{align}
  \sum_{n_{(1,y_2,y_3)}\in \Zb}\exp(in_{(1,y_2,y_3)} W_1(y_2,y_3))=2\pi \delta(W_1(y_2,y_3)).
\end{align}
Hence, the path-integral vanishes unless $W_1(y_2,y_3)=0 \mod 2\pi$. Since the cut-off $a$ is arbitrarily small, we can conclude the path-integral vanishes unless $w_1(x_2,x_3)=0$ up to the gauge transformation.

By the same argument, we can show that the partition function vanishes unless $\int dx_2 A_{\xyz}=\int dx_3 A_{\xyz}=0$ up to the gauge transformation.

So far we have shown that the path-integral of the left-hand side of \eqref{keyidentity} vanishes unless all the Wilson loops vanish:
\begin{align}
  \int dx^4 A_4=\int dx^5 A_5=\int dx^i A_{\xyz}=0,\ (i=1,2,3),\ (\text{up to gauge transformation}).
  \label{vanishingWilsonloops}
\end{align}

\subsubsection{Pure gauge}
Here, we will show that \eqref{vanishingWilsonloops} implies $A_{a}$ is a pure gauge:
\begin{align}
  \exists \lambda(x),\quad A_{a}=\del_{a}\lambda.
\end{align}
Actually, $\lambda$ can be chosen as follows:
\begin{align}
  \lambda(x)=
  \int_{C} \sum_{i=4,5}dy_{i}A_{i}(x_{\xyz},y_4,y_5)+
  \int_{0}^{x_1}dy_1\int_{0}^{x_2}dy_2\int_{0}^{x_3}dy_3 A_{\xyz}(y_{\xyz},0,0),
\end{align}
where $C$ is a path on $T^2_{45}$ starting at $(y_4,y_5)=(0,0)$ and ending at $(y_4,y_5)=(x_4,x_5)$. 

So far, we have shown that Eq.~\eqref{keyidentity} holds up to normalization.
\begin{align}
  \int DB \exp\left(
    \frac{i}{2\pi}\int d^5 x 
    \epsilon^{abc}B_{a}\del_{b}A_{c}
  \right)
  =\mathcal{N}\delta[A],
 \label{keyidentity2}
\end{align}
for a number $\mathcal{N}$. 

\subsubsection{Triviality of the Hilbert space}

Next, we would like to show that $\mathcal{N}=1$ in \eqref{keyidentity2} and complete the derivation of \eqref{keyidentity}.

$\Ncal$ is expressed as
\begin{align}
  \mathcal{N}=\int DADB\exp\left(
    \frac{i}{2\pi}\int d^5 x 
    \epsilon^{abc}B_{a}\del_{b}A_{c}
  \right).
\end{align}
In other words, $\Ncal$ is the partition function of the theory with the dynamical fields $A_a, B_a$ and the action $S=-\frac{i}{2\pi}\int d^5 x 
\epsilon^{abc}B_{a}\del_{b}A_{c}$. 
It is convenient that choose $x_5$ as the Euclidean time direction and consider the operator formalism. Then $\Ncal$ is written in terms of the Hamiltonian $H$ of our theory in the space $T^{3}_{\xyz}\times S^1$ as
\begin{align}
  \Ncal=\Tr[e^{-\ell_5 H}].
\end{align}
We will show the following two facts:
\begin{itemize}
  \item $H=0$.
  \item The Hilbert space is one-dimensional.
\end{itemize}
These two facts imply $\Ncal=1$ and complete the proof of \eqref{keyidentity}.

Only in this subsection, we use Lorentzian action, by introducing the time coordinate $x^0=-ix_5$.  The Lorentzian action is written as
\begin{align}
  S_{L}=\frac{1}{2\pi}\int d^5 x \epsilon^{abc}B_{a}\del_{b} A_{c},\quad a,b,c = 0,\xyz,4,\quad
  \epsilon^{0\xyz 4}=+1.
  \label{LorentzianAction}
\end{align}
Starting from this action, we canonically quantize this system and look at the Hilbert space and the Hamiltonian.

As the first step, we choose the temporal gauge:
\begin{align}
  A_{0}=B_{0}=0.
\end{align}
Then, we have to impose the Gauss law constraints:
\begin{align}
  \del_{\xyz}B_{4}-\del_{4}B_{\xyz}=0,\quad
  \del_{\xyz}A_{4}-\del_{4}A_{\xyz}=0.
  \label{Gausslaw}
\end{align}
There is a residual gauge symmetry in which the transformation parameter is independent of $x^0$.
In this gauge, the action is written as
\begin{align}
  S_{L}=\frac{1}{2\pi}\int d^5 x \left[
    -B_{\xyz}\del_{0}A_{4}
    -A_{\xyz}\del_{0}B_{4}
  \right].\label{LorentzianAction2}
\end{align}
The equations of motion derived from this action imply all the remaining fields are time independent:
\begin{align}
  \del_{0}B_{\xyz}=\del_{0}B_{4}=
  \del_{0}A_{\xyz}=\del_{0}A_{4}=0.
\end{align}

Let us consider further gauge fixing.  Let $w_4$ be the Wilson loop along the $x_4$ direction:
\begin{align}
  w_4(x_{\xyz})=\int dx_4 B_4.
\end{align}
By a residual gauge transformation
\begin{align}
  B_4 \to B_4+\del_4 \lambda,\quad
  \lambda(x_{\xyz},x_4)
  :=-\int_{0}^{x_4} dy_4\left(B_4(x_{\xyz},y_4)-\frac{w_4(x_{\xyz})}{\ell_4}\right),
\end{align}
we can impose the condition
\begin{align}
  \del_{4}B_4=0.
\end{align}
This equation and the Gauss law constraints \eqref{Gausslaw} imply
\begin{align}
  \del_{4}^2 B_{\xyz}=0.
\end{align}
Thus, $\del_{4} B_{\xyz}$ is independent of $x_4$. Moreover, since $\int dx_4 \del_{4} B_{\xyz}=0$ is satisfied, we can conclude
\begin{align}
  \del_{4} B_{\xyz}=\del_{\xyz}B_{4}=0,
\end{align}
in this gauge.
We apply the same gauge fixing for $A_a$.  Then the action \eqref{LorentzianAction2} can be written as
\begin{align}
  S_{L}=\int dx^0 (L_1+L_2),\\
  L_1=\frac{1}{2\pi}\int d^3 x_{\xyz}
  (-B_{\xyz}\del_{0}u_{4}),\label{L1}\\
  L_2=\frac{1}{2\pi}\int d^3 x_{\xyz}
  (-A_{\xyz}\del_{0}w_{4}),\\
  u_4:=\int dx_4 A_4.
\end{align}
Here, $L_1$ and $L_2$ are two copies of an identical theory, and therefore we focus on $L_1$ below.

Now, $B_{\xyz}$ is a function of $x_{\xyz}$ with possible non-trivial Wilson loops. A configuration of $B_{\xyz}$ with an arbitrary non-trivial Wilson loops is given by a linear combination of $\del_{\xyz}f(x_{\xyz};y_{\xyz})$ with $f$ in Eq.~\eqref{unitwindingphi}.  Therefore, the general $B_{\xyz}$ configuration in this gauge can be written in terms of coefficients $v(y_{\xyz})$ as
\begin{align}
  B_{\xyz}(x)=-\sum_{y_{\xyz}\in \Omega}\frac{1}{2\pi}v(y_{\xyz})\del_{\xyz}f(x_{\xyz};y_{\xyz})+b(x_{\xyz}).
  \label{expBxyz}
\end{align}
Here $b(x_{\xyz})$ is a configuration with trivial Wilson loops:
\begin{align}
  \int dx_i b(x_{\xyz})=0,\quad (i=1,2,3).
\end{align}
Such $b(x_{\xyz})$ is eliminated by a gauge transformation.  We also use the cutoff explained in subsection \ref{sec:subsystemsymmetry}, and consider only the Wilson loops at lattice sites $y_{\xyz}\in \Omega$.

On the other hand, since $\del_{\xyz}v_{4}=0$, $v_4$ can be written as
\begin{align}
v_{4}(x_{\xyz})=v_{4,1}(x_2,x_3)+v_{4,2}(x_3,x_1)+v_{4,3}(x_1,x_2).
\end{align}
Then we substitute \eqref{expBxyz} to \eqref{L1} and obtain
\begin{align}
  L_{1}=\sum_{y_{\xyz}\in \Omega}\frac{1}{2\pi}v(y_{\xyz})\del_{0}u_{4}(y_{\xyz}).
\end{align}
The Hilbert space of this theory is the tensor product of many copies of the theory
\begin{align}
  L_{\text{trivial}}=\frac{1}{2\pi}p\del_{0}q,\quad p\sim p+2\pi,\ q\sim q+2\pi.
\end{align}
Each of these theories has a one-dimensional Hilbert space and a vanishing Hamiltonian. Thus, the total Hilbert space, which is the tensor product of all these Hilbert spaces, is one-dimensional.  Also, the total Hamiltonian vanishes.

We completed the derivation of \eqref{keyidentity}.  In this paper, we introduce the cutoff explained in subsection \eqref{sec:subsystemsymmetry}, and show the relation \eqref{keyidentity}.  
Since the lattice spacing $a$ is arbitrary, the identity with the cutoff is enough for the purpose in this paper.

\subsubsection{Exotic Chern-Simons theory}\label{sec:CS}
Let us consider the level one exotic Chern-Simons theory given by the action
\begin{align}
  S_{CS}=\frac{1}{4\pi}\int d^5x \epsilon^{abc}A_{a}\del_{b}A_{c},\quad
  a,b,c=0,\xyz,4.
\end{align}
We can show that this theory also has a one-dimensional Hilbert space.
Actually, by the same procedure as above, we can rewrite the exotic Chern-Simons action as
\begin{align}
  S_{CS}=\int dx^{0} L_{CS},\quad
  L_{CS}=\sum_{y_{\xyz}\in \Omega}\frac{1}{2\pi}v(y_{\xyz})\del_{0}w_{4}(y_{\xyz}),\\
  w_{4}:=\int dx^{4}A_{4},\quad
  A_{\xyz}(x)=-\sum_{y_{\xyz}\in \Omega}\frac{1}{2\pi}v(y_{\xyz})\del_{\xyz}f(x_{\xyz};y_{\xyz}).
\end{align}
Again, the Hilbert space is the tensor product of copies of the one-dimensional Hilbert spaces of the trivial theory.

In this paper, we only consider a torus geometry.  Therefore, the partition function is unity without non-trivial phase.  However, the usual level one Chern-Simons theory is not completely trivial theory, but it is in a non-trivial symmetry protected topological (SPT) phase \cite{Witten:2003ya,Bhardwaj:2020ymp}.  One may expect that our level one exotic Chern-Simons theory is also in a non-trivial SPT phase\footnote{If we impose the magnetic center symmetry, this theory is in a subsystem symmetry protected topological phase as pointed out in \cite{Yamaguchi:2021xeq}.  However, ordinary level one Chern-Simons theory is still in an SPT phase protected by the gravitational symmetry, even if the magnetic center symmetry is not imposed.  It should be interesting to study whether such a purely gravitational SPT phase exists in exotic theories by putting the system on a ``curved'' space-time.}.  In order to see this, one should calculate the partition function of this theory in a more complicated geometry without reflection symmetry.  It will be a nice future problem.

By the way, for the exotic Chern-Simons of level $N$, the number of the dimensions of the Hilbert space is 
$|N|^{\ell_1\ell_2\ell_3/a^3}$.  The exponent is the number of sites $|\Omega|$.  This huge vacuum degeneracy is a typical phenomenon in such an exotic system.

\section{S and T operations}
\label{sec:ST}
We introduce S and T operations in this section, and show that they form the group \slz.  This derivation is parallel to \cite{Witten:2003ya}.

Let us consider theory $Q$ with U(1) subsystem symmetry.  Let the partition function of $Q$ with the background gauge field be
\begin{align}
  Z_{Q}[A]=\int D\phi e^{-I_{Q}[\phi,A]},
\end{align}
where $\phi$ is the dynamical fields of $Q$ and $I_{Q}$ is the action of $Q$.  

We introduce the following shorthand notation for U(1) subsystem gauge fields $A,B$:
\begin{align}
  (A,B):=\frac{i}{2\pi}\int d^5 x \epsilon^{abc} A_{a}\del_{b}B_{c}.
\end{align}
Notice that this pairing is symmetric and bilinear.  Eq.~\eqref{keyidentity} is expressed in this notation as
$\int DB \exp(B,A)=\delta[A]$.

The S operation make a new theory from $Q$ with the background gauge field by 
\begin{align}
  Z_{SQ}[B]=\int DA Z_{Q}[A]\exp(A,B).
\end{align}
The original U(1) subsystem symmetry of $Q$ is gauged by this S operation and is not a global U(1) subsystem symmetry anymore in $SQ$.  Instead, the magnetic center symmetry \eqref{magneticcentersymmetry} is the U(1) subsystem symmetry of $SQ$.

On the other hand, the T operation is defined as
\begin{align}
  Z_{TQ}[A]:=Z_{Q}[A]\exp\frac12 (A,A).
\end{align}
In other words, this T operation just changes the correlation function of currents only by the contact terms.

Let us see these $S,T$ operations generate \slz.  First, let us consider $S^2$:
\begin{align}
  Z_{S^2Q}[C]
  =\int DB\int DA Z_{Q}[A]
  \exp\left(
    (C,B)+(B,A)
  \right)
  =\int DA Z_{Q}[A]\delta(A+C)
  =Z_{Q}[-C].
\end{align}
Therefore, $S^2$ operation is just negating the gauge field or the charge conjugation.  This $S^2$ commutes with $S$ and $T$, and therefore the center of the group.  Also, $S^4=1$ is satisfied.

Next, let us check $(ST)^3$:
\begin{align}
  Z_{(ST)^3Q}[D]=
  \int DADBDC Z_{Q}[A] \exp\left(
    (D,C)+\frac12 (C,C)+(C,B)+\frac12 (B,B)+(B,A)+\frac12(A,A)
  \right)\label{ST^3}
\end{align}
We evaluate $C$ integration first and obtain
\begin{align}
  \int DC 
  &\exp\left(
    (D,C)+\frac12 (C,C)+(C,B)
  \right)\nonumber\\
  \qquad &=\int DC 
  \exp\left(
    \frac12 (C+B+D,C+B+D)-\frac12(B+D,B+D)
  \right)\nonumber\\
  &=Y
  \exp\left(
    -\frac12(B+D,B+D)
  \right),\\
  Y&=\int DC \exp \frac12 (C,C).
\end{align}
Here $Y$ is just a numerical factor independent of the background gauge fields.  Actually in our spacetime $T^3_{\xyz}\times T^2$, $Y=1$ since the Hamiltonian vanishes and the Hilbert space is one dimensional as shown in subsection \ref{sec:CS}.
In general, we expect $Y$ is an invertible phase.

Substituting this result into Eq.~\eqref{ST^3}, we obtain
\begin{align}
  Z_{(ST)^3Q}[D]
  &=Y
  \int DADB Z_{Q}[A] \exp\left(
    -\frac12(B+D,B+D)
    +\frac12 (B,B)+(B,A)+\frac12(A,A)
  \right)\nonumber\\
  &=Y
  \int DA Z_{Q}[A] \exp\left(
    -\frac12(D,D)
    +\frac12(A,A)
  \right)\delta(D-A)\nonumber\\
  &=YZ_{Q}[D].
\end{align}
Since $Y=1$ in our setup, we conclude $(ST)^3=1$, and therefore $S,T$ generate \slz.

Although we focus on the partition function in this paper, \slz transformation defines not only partition functions, but local quantum field theories coupling to the background gauge field.  For example, the dynamical fields of $SQ$ are $\phi$ and $A$.  The action of $SQ$ with the background field $B$ is given by
\begin{align}
  I_{SQ}(\phi,A,B)=I_{\phi}(\phi,A)-(A,B).
\end{align}

\section{\texorpdfstring{$\phi$}{phi} theory}
\label{sec:phi}
In this section, we take the $\phi$ theory \cite{Seiberg:2020bhn,Gorantla:2020xap} as an example and consider S and T transformation of this theory.

\subsection{Definition and partition function}
Let us consider a bosonic field $\phi$ that has periodicity $\phi\sim \phi+2\pi$. The action is written as
\begin{align}
  \Iphi[\phi,0]=\int d^5 x \Lcal_0,\quad
  \Lcal_0=\frac{\mu_0}{2}\del_{a}\phi \del^{a}\phi.
\end{align}

The $\phi$ theory has U(1) subsystem symmetry: the shift of $\phi$.  We can couple this U(1) subsystem symmetry to the background gauge field:
\begin{align}
  \Iphi[\phi,A]=\int d^5 x \Lcal,\quad \Lcal=\frac{\mu_0}{2}D_{a}\phi D^{a}\phi,\quad D_{a}\phi:=\del_{a}\phi-A_{a}.
  \label{gaugedphitheoryaction}
\end{align}
This theory is invariant under the following gauge transformation
\begin{align}
  \phi\to \phi'=\phi+\lambda,\qquad
  A_{a}\to A'_{a}=A_{a}+\del_a \lambda,
\end{align}
where $\lambda$ is the parameter that has the same properties as the $\phi$ field.  The partition function of this theory is given by
\begin{align}
  Z_{\phi}[A]=\int D\phi \exp(-\Iphi[\phi,A]). \label{Zphi}
\end{align}

Since the above theory is a free theory, we can exactly calculate the partition function \eqref{Zphi}.
To do this, we consider the Fourier transformation\footnote{In this procedure, we take the large volume limit and ignore the winding modes.  In this sense, our calculation in this section is only perturbative.}
\begin{align}
  \phi(x)=\int_{k}\phit(k)e^{ik\cdot x},\quad
  A_{a}(x)=\int_{k}\At_{a}(k)e^{ik\cdot x},
\end{align}
where $k$ is a 5-vector $k^{\mu},\ \mu=1,2,3,4,5$.  In addition, we use the following shorthand notation:
\begin{align}
  k\cdot x:=k_{\mu}x^{\mu},\quad \int_{k}:=\int \frac{d^5 k}{(2\pi)^5}.
\end{align}
It is convenient to define $k_{\xyz}$ as
\begin{align}
  k_{\xyz}:=k^{\xyz}:=-k_1k_2k_3.
\end{align}
Then, we can express
\begin{align}
  \del_{a}\phi(x)=\int_{k}\phit(k) ik_a e^{ik\cdot x},\ (a=\xyz,4,5).
\end{align}

We can evaluate the Gaussian integral of \eqref{Zphi} and obtain
\begin{align}
  Z_{\phi}[A]
  &=Z[0]\exp\left(
    -\frac{\mu_0}{2}\int_{k}\frac{\delta_{ab}k^2-k_ak_b}{k^2}\At_{a}(k)\At_{b}(-k)
  \right)\nonumber\\
  &=Z[0]\exp\left(
    -\frac{\mu_0}{4}\int d^5x d^5 y F_{ab}(x)F_{ab}(y)G(x-y)
  \right),\label{partitionfunctionphi}\\
  F_{ab}&:=\del_a A_b -\del_b A_a,\quad
  G(x-y):=\int_{k}\frac{1}{k^2} e^{ik\cdot (x-y)},\quad   k^2:=k_ak^{a}.
\end{align}
Since $F_{ab}$ is gauge invariant, the partition function is explicitly gauge invariant.

\subsection{\slz transformation}
In this subsection, we consider \slz transformations of $\phi$ theory.

Let us first consider S transformation of the $\phi$ theory.  The S transformation is making the gauge field dynamical in \eqref{gaugedphitheoryaction}, so we expect the S transformed theory is an ``empty'' theory.

The partition function of S transformed theory with the background gauge field $B$ is
\begin{align}
  Z_{S\phi}[B]
  &=\int DA Z[A]\exp\left(\frac{i}{2\pi}\int d^5 x e^{abc}B_{a}\del_{b}A_{c}\right).
\end{align}
After gauge fixing and ignoring decoupled ghosts (see appendix \ref{app:gaugefixing})
\begin{align}
  Z_{S\phi}[B]&=\int DA \exp\left(-\frac12 \int_{k}\left[\mu_0 \delta_{ab}\At_{a}(k)\At_{b}(-k)
  -\frac{2}{2\pi}\epsilon^{acb}k_c\Bt_{a}(k)\At_{b}(-k)\right]\right).\label{SZ}
\end{align}
We can evaluate the Gaussian integral in \eqref{SZ} and obtain
\begin{align}
  Z_{S\phi}[B]
  =&Z_{S\phi}[0] \exp\Bigg[-\frac{1}{2(2\pi)^2\mu_0} \int_{k}(k^2\delta_{ab}-k_{a}k_{b})\Bt_{a}(k)\Bt_{b}(-k)
  \Bigg]\notag\\
  =&Z_{S\phi}[0] \exp\Bigg[-\frac{1}{4(2\pi)^2\mu_0} \int d^5 x G_{ab}G_{ab}
  \Bigg],\label{ZSphi}\\
  G_{ab}:=&\del_a B_b-\del_b B_a.\notag
\end{align}
Since the quadratic term is local, the 2-point functions of currents have only contact terms.  Actually, two point function is given by
\begin{align}
  \left\langle J^a(x) J^b(y)
    \right\rangle
    =\frac{1}{Z_{S\phi}[0]}\frac{\delta}{\delta B^{a}(x)}\frac{\delta}{\delta B^{b}(y)}Z_{S\phi}[B] \Bigg|_{B=0}
    =\frac{1}{g^2}(\delta^{ab}\del_{c}\del_{c}-\del_{a}\del_{b})\delta^5(x-y).
\end{align}
In particular, $\left\langle J^a(x) J^b(y)
\right\rangle=0$ if $x\ne y$.
This is consistent with the fact that the S transformed theory is ``empty'' and contains no propagating degrees of freedom as we expected.

Second, let us consider $S^2$ transformation of the $\phi$ theory.  By the general discussion in section \ref{sec:ST}, $S^2\phi$ is nothing but the $\phi$ theory $\Iphi(\phi, -A)$ of \eqref{gaugedphitheoryaction}. On the other hand, from the explicit calculation \eqref{ZSphi}, the partition function of $S^2\phi$ is given by
\begin{align}
  Z_{S^2\phi}[A]=Z_{S\phi}[0] \int DB\exp\Bigg[-\frac{1}{4(2\pi)^2\mu_0} \int d^5 x G_{ab}G_{ab}+\frac{i}{2\pi}\int d^5 x \epsilon^{abc}A_{a}\del_{b}B_{c}
  \Bigg].
\end{align}
Thus, $S^2\phi$ theory is the exotic Maxwell theory with the coupling constant $2\pi \sqrt{\mu_0}$ and the dynamical Maxwell field $B$ that couples to the background gauge field $A$.  By comparing these two results, we find that the $\phi$ theory is equivalent to the exotic Maxwell theory.

Third, let us consider the T transformation of the $\phi$ theory.  The partition function of $T\phi$ is given by
\begin{align}
  Z_{T\phi}[A]=Z_{\phi}[A]\exp\left(\frac{i}{4\pi}\int d^5 x \epsilon^{abc}A_{a}\del_{b}A_{c}\right),
\end{align}
where $Z_{\phi}[A]$ is the partition function of the $\phi$ theory \eqref{partitionfunctionphi}.

Finally, let us next consider $ST^w$ transformation for an integer $w$.  The Fourier transformation of the Chern-Simons term is
\begin{align}
  &\frac{i}{4\pi}\int d^5 x \epsilon^{abc}A_{a}\del_{b}A_{c}
  =-\frac12 \int_{k} (-K^{ab}(k))\At_{a}(k)\At_{b}(-k),\\
  &K^{ab}(k):=\frac{1}{2\pi} \epsilon^{acb}k_{c}.
\end{align}
Then
\begin{align}
  &Z_{ST^w\phi}[B]=\int DA \exp\left(
    -\frac12 \int_{k}\left[
      M_{ab}(k)\At_{a}(k)\At_{b}(-k)-2K^{ab}\Bt_{a}(k)\At_{b}(-k)
    \right]
  \right),\\
  &M_{ab}(k):=\mu_0 \delta_{ab}-wK_{ab}(k).
\end{align}
As the same calculation as above, we can show
\begin{align}
  Z_{ST^w\phi}[B]=Z_{ST^w\phi}[0]
  \exp\left(
    -\frac12 \int_{k}\frac{1}{(2\pi)^2\mu_0^2+k^2 w^2}\left[
      \mu_{0}(k^2\delta_{ab}-k_ak_b)+
      \frac{w k^2 k_c}{2\pi}\epsilon^{abc}
    \right]\Bt_a(k)\Bt_b(-k)
  \right).
\end{align}
Thus, by this transformation, we can produce infinite series of QFT's with U(1) subsystem symmetry.

\subsection*{Acknowledgement}
The author would like to thank Sinya Aoki, Yui Hayashi,	Masazumi Honda, Hayato Kanno, Masashi Kawahira, Shigeki Sugimoto, Hidehiko Shimada, Ken Shiozaki, Yuya Tanizaki, Seiji Terashima, Masataka Watanabe, Shuichi Yokoyama for useful discussions and comments.  He would also like to thank the Yukawa Institute for Theoretical Physics at Kyoto University for hospitality during his stay. Discussions during the YITP workshop YITP-W-22-09 on ``Strings and Fields 2022'' were useful to complete this work. 
This work was supported in part by JSPS KAKENHI Grant Number 21K03574.

\appendix
\section{Gauge fixing}
\label{app:gaugefixing}
We employ BRST formalism to fix the gauge.
We introduce the ghost fields $c(x),\ b(x),\ B(x)$.  Here $c(x),\ b(x)$ are fermionic fields and $B(x)$ is a bosonic field.  The BRST transformation $\deltaB$ is defined as
\begin{align}
  \deltaB A_a(x)=\del_{a} c(x),\quad
  \deltaB c(x)=0,\ 
  \deltaB b(x)=B(x),\ 
  \deltaB B(x)=0.
\end{align}
Notice that $\deltaB^2=0$.

We add the following BRST exact term to the action
\begin{align}
  S_{\mathrm{GF}}
  =&\deltaB \int d^5x  \frac{\mu_0}{2}\del_{a}b(x)(-\del_{a}B(x)+2A_{a}(x))\\
  =&\int d^5 x \frac{\mu_0}{2}\left[
    \del_{a}B(x)(-\del_{a}B(x)+2A_{a}(x))
    -2\del_{a}b(x)\del_{a}c(x)
  \right].
\end{align}
The $b(x),\ c(x)$ term decouples from the other parts, so we just integrate them out.  We integrate out $B(x)$ field and obtain
\begin{align}
  S_{\mathrm{GF}}
  =&\int d^5 x \frac{\mu_0}{2}\left[
    \del_{a}B(x)(-\del_{a}B(x)+2A_{a}(x))
  \right]\nonumber\\
  =&\frac{\mu_0}{2}\int_{k}\frac{k_{a}k_{b}}{k^2}\At_{a}(k)\At_{b}(-k).
\end{align}

\bibliographystyle{utphys}
\bibliography{ref}
\end{document}